27

# Deep opacity in AI: A treat to XAI and standard privacy protection mechanisms

*Vincent C. Müller*

March 2025[*]

*Abstract:* It is known that big data analytics and AI pose a threat to privacy, and that some of this is due to some kind of "black box problem" in AI. I explain *how* this becomes a problem in the context of justification for judgments and actions. Furthermore, I suggest distinguishing three kinds of opacity: 1) the subjects do not know what the system does ("shallow opacity"), 2) the analysts do not know what the system does ("standard black box opacity"), or 3) the analysts cannot possibly know what the system might do ("deep opacity"). If the agents, data subjects as well as analytics experts, operate under opacity, then these agents cannot provide justifications for judgments that are necessary to protect privacy, e.g., they cannot give "informed consent", or guarantee "anonymity." It follows from these points that agents in big data analytics and AI often cannot make the judgments needed to protect privacy. So I conclude that big data analytics makes the privacy problems worse and the remedies less effective. As a positive note, I provide a brief outlook on technical ways to handle this situation.

---

[*] A preliminary and shorter version of this article was presented at the AISB 2021: Müller, Vincent C. (2021), 'Deep opacity undermines data protection and explainable artificial intelligence', in Carlos Zednik and Hannes Boelsen (eds.), *AISB 2021 Symposium Proceedings: Overcoming Opacity in Machine Learning* (Bath: AISB), 18-21.

## 27.1. Background: Opacity and Data Protection

### 27.1.1. Opacity and Justification

It seems that the literature on opacity or explainable AI has not sufficiently "framed" the issue, i.e., it has not sufficiently explained why "opacity" or the "black box problem" in AI is an issue, when, and for whom. It is not clear that we can assume there is *one* overarching question that can be dealt with through *one* theory. As in any conceptual issue, it is good to state where the interest in the problem originates from. Here we start from an explicitly stated societal problem, data protection, and then show how this leads to a taxonomy of opacity at three levels: shallow, standard, and deep.

I will suggest that at bottom, the phenomenon of epistemic "opacity" in computational modeling (Humphreys 2009), machine learning (Burrell 2016), and data science (Symons and Alvarado 2016) stems from our human practice of explaining and justifying our actions, especially our decisions. This is why we are worried when a computational system just decides something ("Your application for a credit card is rejected"), and does not provide any further justification. This is why we are concerned that such decisions may be based on the wrong criteria, i.e., they may be unfair because they rely on a bias that discriminates against certain groups (e.g., foreigners). When we ask "why?", in such a context, we are not just looking for an *explanation* in *causal* terms, we are looking for a *justification* in terms of *reasons*. In the classic paper, Humphreys had defined "[A] process is epistemically opaque relative to a cognitive agent X at time t just in case X does not know at t all of the epistemically relevant elements of the process" (Humphreys 2009, 618). This was usefully focused on the issue of justification by Durán and Formanek (2018): "[A] process is epistemically opaque relative to a cognitive agent X at time t just in case X at t doesn't have access to and can't survey all of the steps of the justification." So this discussion falls mainly within the realm of normative discourse. It is part of the explanation why we worry about the allocation of "responsibility" in the context of AI, and about the criteria for "moral agency." Opacity is a central issue in what is now often called "data ethics" (Floridi and Taddeo $_{2016}$; Mittelstadt and Floridi 2016).

One may wonder whether a demand for transparency is not asking a bit too much of humans and machines – but what we ask for is not full transparency but justification to an extent. Human practices and demands on the ability to justify actions vary strongly with context – when the stakes are high, they are extremely high and codified, e.g., in legal decisions and in safety-critical contexts, whereas in many other contexts they are more relaxed. It is an open question how far "explicability" should go, given that the abilities of humans in that regard are not too impressive. Zerilli et al. (2019) argue that there is a double standard.

Of course, humans hardly ever know the full implications of their actions, or have full knowledge of all relevant facts. In fact, the "epistemological frame problem" seems to indicate that such knowledge is impossible because it is never clear in advance which knowledge might turn out "relevant" and what needs to be updated after an action (Dennett 1984).

There are several technical activities that aim at "explainable AI," starting with Lomas et al. (2012) and Van Lent et al. (2004) and, more recently, a 2017–21 DARPA research programme (Gunning 2017) and the AI4EU.eu project on "human-centred AI." More broadly, the demand for "a mechanism for elucidating and articulating the power structures, biases, and influences that computational artefacts exercise in society" (Diakopoulos 2015) is sometimes called "algorithmic accountability reporting." This does not mean that we expect an AI to "explain its reasoning" – doing so would require far more serious moral autonomy than we currently attribute to AI systems. What we are currently looking for is to enable humans who use the AI system to explain their decisions which were based on the AI system.

### 27.1.2. Big Data

It turns out that the opacity issue is not limited to AI; it appears in data analytics, also, particularly in big data analytics. In a first approximation, "big data" analytics really is characterized by just being "big," i.e., it is not a radically new technique but it does what has been done elsewhere on a big scale. Having said that, we often think that a significant increase on a gradual scale can make the difference that is worthy of a new name and separate scientific treatment (so we distinguish babies from adults); also, such a significant increase can pass a threshold and lead to entirely different system state (add more weight and the bridge collapses). In the case of big data, the measurement of size can be done in various dimensions, but the classical ones are data at high *volume*, with high *velocity*, and *variety* – big data is often characterized through these "3Vs." And indeed, the 3Vs are changing things in an impressive way.

The volume of data is enormous: 90% of the total digital data extant today was only created in the last 2 years. We add 200 million emails and 200,000 Instagram posts per minute (IBM 2016), but far more data is added automatically by digital processes – for example, each of the 20 sensors on a single GE wind turbine generates 520GB of data per day (Saran 2013). The velocity of this addition is such that many systems stream data continuously in real time, e.g., video cameras and other sensing devices.

Note the importance of technologies like "smart cities," "smart homes" (including "smart light bulbs," "smart floors," "smart doors," etc.), "smart phones," "wearables," the "quantified self" (e.g., 23andMe.com), electronic toys and the "Internet of Things" – these are signs of the general digitalization of the human environment, mainly through more sensors, that automatically generates a massive *volume* of data, with high *velocity* and *variation* (3Vs). So, there is now about nine times more digital data than 2 years ago, not just because almost all the classical data (medical records, business data, governmental data, etc.) is now digital, but because our lives are conducted to a large extent digitally, and there are many more sensors that generate a stream of digital data from the nondigital aspects of the world. This last trend, particularly, is the driving force of the "big" data increase.

### 27.1.3. Big Data and Surveillance

This big data contains valuable information that analysis can reveal; in particular, it can reveal patterns of how certain parts of the data relate to other parts, whether or not the data was generated for this purpose. One example from medical research is that the continuous flow of monitoring data of prematurely born babies in neonatal intensive care (EEG, ECG, blood oxygen, blood pressure, temperature, drug infusion pump data, growth data, etc.) is normally recorded at regular intervals, or used for alarms – i.e., almost all data is discarded. When such data was fed into a big data system under the supervision of medical experts (McGregor 2013), researchers were able to find early signs for the onset of potentially fatal conditions – patterns that indicated something was coming before any pathological symptoms occurred. Following this discovery, real-time analysis is now used to detect these patterns and save lives (Miotto et al. 2016).

Of course, the patterns might also say other interesting things, e.g., whether someone is likely to buy a particular product, what political views they might hold, or what sexual preferences they may have. The detection and analysis of these patterns in ever more data are now part a business model that is used on a large scale in IT services that are provided "free of charge." "Surveillance is the business model of the Internet" (Schneier 2015). Why are they provided without an obvious payment? Because the payment is in the form of data, or, as the saying goes: "If you are not paying for the product then you are the product." So, companies like Google provide their services and connect their use to my ID (email, phone number, name, address), so they know my various devices, email traffic, account use, log-on patterns, search patterns, YouTube use, gaming use, cloud documents, friends and collaborators, photos and videos (geo-tagged), calendar, address book, location, etc., etc. As a PBS documentary sums it up, "In this vast ocean of data, there is a frighteningly complete picture of us" (Smolan 2016, 1).

In some ways, Google knows more about me than any person does; it even knows many things about me that I don't know myself. Perhaps I am more right wing or more homosexual than I would like to admit. And of course, I forget lots of details but Google doesn't. And if I turn off my mobile phone or don't use a service ... well, that's a pattern, too. And that is just one company. Add government sources that may be able to access more than just one company or add classical surveillance from sensing in homes and on the streets (microphones, cameras, motion sensing, temperature, etc.), add private peer-to-peer surveillance and we are already pretty close to a total surveillance situation – without noticing much of it. My personal data has become a commodity that is accessed out of my control.

The ethical evaluation of this situation may just turn on the consequences and in this situation there is a (small) probability that the overall outcome may be more beneficial than a scenario in which it is transparent who has access to my personal data and where I can control that access. However, the current situation is problematic from a rights perspective since it involves deception and, crucially, it does not respect a right to privacy, in particular informational privacy, which is usually considered to be fundamental for a person's autonomy (Roessler 2017). The question is what can be done to protect and continue positive use of AI and data analytics, while protecting privacy. We thus need a brief look at the requirements for policy in the area.

### 27.1.4. Standard Regulations

In the EU, many of these issues have been taken into account with the General Data Protection Regulation (GDPR 2016). The GDPR was agreed in the European Parliament and the EU Council in April 2016 and became law from May 2018 in the member states. Member states have the right to specify further rules that do not contradict this regulation (Section 10, cf. Section 13). It is a very powerful regulation that embeds a number of new features, while being the result of extended negotiations. I expect it to be a setting the political debate for a number of years and would be surprised if substantial changes in the political and regulatory situation were to occur in the next 10 years. Note that the major prior EU document, 95/46/EC, was approved 21 years earlier, only at the level of a "directive" for national law (EU Parliament 1995), whereas the GDPR automatically became law itself. The GDPR foresees a "right to explanation" – although the extent to which this goes and to which it can be enforced is disputed (Goodman and Flaxman 2017; Wachter et al. 2017, 2018). In any case, an inability to explain decisions appears to violate due process, especially when such decisions are challenged.

I will thus take this regulation as a general indicator of what kind of criteria are needed to act lawfully and, to some extent, ethically in privacy matters. It is not my aim to evaluate this regulation, but rather to take it as "what we've got" and see what kinds of problems we can expect in our special case: big data.

The GDPR can be summarized in the following points.
1. It concerns "personal data": name, address, localization, online identifier, health information, income, cultural profile.
2. Communication: who gets the data, why, for how long? (No use for other "incompatible" purposes. Use as long as necessary.)
3. Consent: get clear informed consent.
4. Access: provide access to my data.
5. Right to be forgotten (not for research).
6. Right to explanation for contracts (and right to have a person decide).
7. Marketing: right to opt out.
8. Legal: maintain EU legislation when transferring data out.
9. Need for a "data protection officer" in your organization.
10. Impact assessment prior to high-risk processing (new technology, personal information, surveillance, sensitive).

The crucial points for our discussion of opacity are no. 3 (informed consent) and 6 (right to explanation), to a lesser extent 4 (access to my data) and 5 (right to be forgotten). To stress this again, these demands of an exemplary data protection regulation like the GDPR just reflect the demands that we make on human agents: to be able to justify their decisions and actions to some extent.

## 27.2. Types of Opacity

Here, we will try to explain opacity as a property of the relation between a person and a system.

### 27.2.1. Shallow Opacity: AI as an Instrument of Power

AI is used in automated decisions systems and decision support systems, especially through "predictive analytics." The output of such a system may be relatively trivial, such as "this restaurant matches your preferences," "the patient in this X-ray has completed bone growth," or have greater significance for a person, e.g., if it says "application for credit card declined," "donor organ will be given to another patient" or "target identified and engaged." At the same time, it will often be impossible for the affected person to know how the system reached this output, i.e., the system is "opaque" to that person, who is typically also the/a data subject. The institution using the system (e.g., the bank) may just tell me that "the system has decided" but not how and why – even if the institution can find out what the reasons were.

In such cases, opacity is just a matter of decision from some party that is in power and which could find a justification, if it wanted. The opacity in these cases I shall call "shallow opacity." This kind of asymmetric opacity to *users only* exemplifies the classic (and important) slogan "knowledge is power." The users have no control over output and thus are not responsible for the output. This kind of opacity is not specific to AI – it can happen in any use of data science for a decision or decision support system.

This has been used to classify the notion of opacity in general: "They are opaque in the sense that if one is a recipient of the output of the algorithm (the classification decision), rarely does one have any concrete sense of how or why a particular classification has been arrived at from inputs" (Burrell 2016, 1), but as we shall see, it is really a special case. It appears that shallow opacity is the notion Surden and Williams (2016) had in mind when they talk about "technological opacity": "'technological opacity' applies any time a technological system engages in behaviors that, while appropriate, may be hard to understand or predict, from the perspective of human users."

It is often said that such matters raise "significant concerns about lack of due process, accountability, community engagement, and auditing" (Campolo et al. 2017, 18ff). These algorithmic systems are part of a power structure, which is why Danaher talks about an "algocracy" and concludes that "we are creating decision-making processes that constrain and limit opportunities for human participation" (Danaher 2016, 245). Again, however, the issue of power structure also applies when the opacity does not concern only the user or data subject – this kind of opacity is the subject of our next section.

### 27.2.2. Standard or "Black Box" Opacity

Many AI systems rely on machine learning techniques in (simulated) neural networks that will extract patterns from a given dataset through "learning." These networks are organized in "layers," one of which is the "input layer" and one the "output layer, with one or many "hidden" layers in between. If there is more than one such hidden layer, the network is often called a "deep" neural network (DNN) and the learning is "deep learning" (Goodfellow et al. 2016). Data either flows from one layer to the next in one direction, in "feed-forward" systems, or in any direction, in "recurrent" systems. The network can be recalibrated through a feedback system, which changes the outcome on a given income, i.e., the system "learns."

These networks broadly learn in three different ways: supervised, semi-supervised (e.g., reinforcement) or unsupervised – though these ways are not mutually exclusive. In the "supervised" case, the system is advised whether an output is correct or incorrect, or is shown "good" outputs (AlphaGo was of this sort, which beat a top-ranked Go player in 2016, using a supercomputer with 1920 processors and 280 GPUs) (Economist 2016). In the "reinforcement" case, it is told about a broader target (e.g., win the game of Go) but not whether an output (e.g., a move) was correct or incorrect (AlphaGo Zero is of this sort) (Silver et al. 2018). Finally, in the unsupervised case the system tries to find patterns by itself, without feedback on which patterns are useful or correct (these systems are of central importance in statistics) (Hinton and Sejnowski 1999). We may find patterns we were not looking for, that nobody knew, and that we cannot explain.

With these techniques, the "learning" captures patterns in the data and these are labeled in a way that appears useful to the programmer while the programmer does not really know how these patterns came about: "We can build these models, but we don't know how they work" (Knight 2017). In fact, the programs are typically evolving so when new data comes in or new feedback is given, the patterns in the learning system change. There is a significant recent literature about the limitations of machine learning systems (Castelvecchi 2016; Marcus 2018), that are essentially sophisticated data filters – and quite possibly at the peak of the "hype cycle" at the moment. Furthermore, the quality of the program depends heavily on the quality of the data provided, following the old slogan "garbage in, garbage out." So, if the data already involved a bias (e.g., police data about the skin color of suspects or job data including gender), then the program will reproduce that bias. There are proposals for a standard description of datasets in a "datasheet" that would make the identification of such bias more feasible (Gebru et al. 2018).

What this means for our purposes is that the outcome cannot really be explained, it is opaque to the user or programmers, especially but not uniquely in the less supervised learning ways. It is thus more opaque than the cases of "shallow opacity" above, in that the opacity does not just apply to the users or data subjects; it also applies to the experts – the agent relativity was stressed by (Zednik 2019). Opacity for experts is not only about what the experts know at a particular point in time, but also about what they can know, even after research. For that reason, the resulting AI is often called "black box AI" – it features a black box between input and output rather like the human mind is a black box in the eyes of a (methodological) behaviorist.

This kind of opacity is what is often mentioned in general discussions of opacity, so I call it "standard opacity." It is, however, specific to AI, in fact to a particular method of AI, namely machine learning. It features the problems of distribution of power mentioned under shallow opacity, and it shows the inability to provide justification for its output. It is probably what authors have in mind when they say that the systems "… are opaque: it is difficult to know why they do what they do or how they work," how to explain their "explanatory success" (Zednik 2019). The systems know things but we do not, i.e., in the terminology of the "Rumsfeld cases" there are "unknown knowns" (Nickel 2019).

Perhaps the issue of democratic legitimacy is more urgent in the case of standard opacity, since it cannot be easily relieved. Kissinger pointed out that there is a fundamental problem for democratic decision making if we rely on a system that is supposedly superior to mere humans, but cannot explain its decisions. He says we may have "generated a potentially dominating technology in search of a guiding philosophy "(Kissinger 2018). In a similar vein, Cave (2019) stresses that we need a broader societal move toward more "democratic" decision making to avoid AI being a force that leads to a Kafka-style impenetrable suppression system in public administration and elsewhere.

### 27.2.3. Deep Opacity

What has not been sufficiently taken into account in the discussions of opacity and the "black box" is *to whom* the system is opaque, and *to what extent*. It can be opaque to a user but not to the programmer. It may be opaque to both, but in a way that the expert analyst can overcome. I suggest there is a third case, where opacity cannot possibly be removed, even for the human expert, even if best efforts are made in the generation of the algorithms; this I call "deep opacity."

In order to introduce deep opacity, it will be useful to remind ourselves of the kinds of questions we are supposed to answer, where opacity gets in the way – this follows from the section "Standard regulations" above. We said "The crucial points for our discussion of opacity are no. 3 (informed consent) and 6 (right to explanation), to a lesser extent 4 (access to my data) and 5 (right to be forgotten)." Some questions that follow from these demands are listed below.

- Does this data include information about me?
- Can you give me access to all the data about me?
- Does this data include personal information?
- Does this data include a particular piece of information "that p"?
- Is the data in this dataset anonymous? Can it be deanonymized?
- What information can be derived from this data?

Data analytics and AI for data are ways to find information that is in the data. Prior to carrying out the analysis, and prior to combining this data with other data, we cannot even know what kind of result the analysis will reveal. It might even reveal patterns that are unknown to the subject of the data itself, if the data is about a person or a company. It will reveal statistical patterns that only allow predictions with a certain degree of certainty – and correlation is not causation. It can be shown formally that certain sets of queries to a database will reveal the entire database (Abowd 2017) or that the data can be reidentified. No matter how well blended the data soup is, there are ways to de-blend some of it and find some information.

Anonymization is a case in point. A dataset that provably cannot reveal any information about a particular item in it would have to be devoid of information. If there is information, who knows how useful it can be? For example, the nationality of individuals may reveal very little but if a particular piece of data is a rare property, perhaps even unique, it can unravel the whole information about that person. There is a host of examples of databases that were deemed safe to be released into the public domain – and then a smart way to deanonymize them was found (Rocher et al. 2019).

The data contains information, but how much? We cannot know. We can only know what a particular method can reveal on a particular dataset, not what future methods might reveal or what a dataset might reveal when combined with another dataset. So even under ideal conditions and with perfect expert knowledge, we cannot justify answers to the questions above. That is deep opacity.

### 27.3. Ethical Judgment under Opacity

If an agent operates under opacity (on a set of data), they cannot judge certain things, e.g., whether that data constitutes a threat to privacy. They can also not "give informed consent" to the use of data since being "informed" would require knowing at least the main implications of giving that consent. I cannot be responsible for what I cannot know (which is more than what I can know, which is more than what I do know).

There is a long tradition in ethics that recognizes what one knows as an important factor in how one's actions are evaluated. This tradition is especially important and developed in criminal law, so I will use this standard to explain – note that the terminologies used differ in different jurisdictions.

Say you perform an action that results in the death of another human. Let us assume you acted without legal justification, so not as a solider or hangman, for example; not in self-defence and not on demand of the victim. Whether you are guilty of a crime, and which one, depends on what your intentions were, what you believed at the time, and what you should have known at the time. So, for example, in 2013 the well-known runner Oscar Pistorius shot his girlfriend Reeva Steenkamp. Pistorius admitted shooting her. He also admitted intentionally shooting his gun. However, he shot her at night, through a closed toilet cubicle door and claimed he had believed a burglar to be in that room – so he claimed not to have intentionally shot *her*, in particular. The sentencing depended on what he believed at the time and what he should have known but perhaps was negligent of, For these reasons, his defence argued for dismissal due to "self-defence" but Pistorius was sentenced for "culpable homicide" due to negligence in 2014 (Lyall 2014), and then in a retrial for "murder" (due to intent to kill whoever was in the bathroom) in 2016; appeals by both the prosecution and the defense were rejected (Onishi 2016). These discussions on the role of ignorance in ethics go back to Thompson (1976); for more recent developments, see Peels (2017).

Opacity in big data analytics is thus truly a dilemma. Big data analytics always has deeper opacity (it is definitional for this field), and if we have deeper opacity, then we do not have standard data protection. So, with big data we get the worst of both worlds: increasing potential to do harm, with less ability to find out what is ethically right and to enforce what we think is right.

### 27.4.    Outlook: Data Science under Opacity

Given that we value data protection and data analytics, we must try to strike a balance between these two – or determine some absolute rules that must be respected, even if violating these rules would increase overall utility. For a data scientist, the situation presents itself as follows: "It is easy to say that anonymization is impossible and that re-identification can always take place. It is just as easy to be complacent about the privacy risk posed by the availability of anonymized data. It is more difficult to evaluate risk realistically and in the round and to strike a publicly acceptable balance between access to information and personal privacy" (Elliot et al. 2016, vi).

There are some promising ways to avoid these problems, following the slogan "as much data as necessary." Most uses of data science aim for general statistical insight, not for insight about particular individuals. So, for example, an epidemiologist who works on the causal factors of a particular disease does not need data about individuals but only about populations – she can work with "aggregated data." In order to prevent deanonymization, this data can be "stuffed" with additional synthetic data points, while preserving the informational content that is relevant for the particular research. Such data can also be used for training machine learning systems. The conclusions from such data can be used to explain individual cases, though this is known to raise significant problems (Garrett 2003).

This idea is often called "differential privacy" where there is a guarantee that no method can lead from publicly available output back to individual private data; in other words, any combination of queries to a database should not reveal private data. This is achieved by adding calibrated "noise" to the output to queries (Dwork et al. 2006). It is also possible to add this noise at the source of data collection, so that identifiable data is never even generated, or transmitted. One way to achieve this is as follows – a randomly selected (e.g., via the toss of a coin) half of the respondents answer randomly, the other half answer truthfully. This version is called "local differential privacy" because it produces the desired data (if the sample is large enough) but no "local" dataset with personal data is ever generated (Abowd 2017). Local differential privacy is apparently used by Apple to generate user data on iOS while guaranteeing that the company does not know what an individual user has done (Apple Differential Privacy Team 2017; Greenberg 2016).

## 27.5. Conclusion

We have proposed an analysis of opacity in AI and data science that is shaped by the societal context in which the issue arises, namely data protection and justification in automated decision systems. It appeared that there are three main types of opacity: "shallow" opacity, which is a matter of power structures; "standard" opacity, which is a matter of processing method; and "deep" opacity, which is a matter of informational content and robustness in the face of new methods or additional data. It would appear that avoiding the standard and deep varieties of opacity would require massive changes in the practices of data science, in particular the avoidance of identifiable data altogether, as well as openness about data sources, processes, and stakeholders – this is not impossible, but it may prove a large demand for this very fruitful new science.